\documentclass[]{easychair}

\usepackage{relsize}
\usepackage{tikz}
\usetikzlibrary{arrows,matrix,automata,shapes,snakes}
\usepackage{amssymb}
\usepackage{amsmath}
\usepackage{doc}
\usepackage{makeidx}
\usepackage{color}
\usepackage{colortbl}
\usepackage{graphicx}
\usepackage{float}
\usepackage{subfig}
\usepackage{xspace}

\newcommand\cv{2.0\xspace}

\def\systemname#1{\textsf{#1}\xspace}
\newcommand{\HOLLight}{\systemname{HOL Light}}
\newcommand{\KBCV}{\systemname{KBCV}}
\newcommand{\MKBTT}{\systemname{MKB{\scalebox{0.9}{TT}}}}

\newcommand{\TTTT}{%
 \systemname{T\kern-0.2em\raisebox{-0.3em}T\kern-0.2emT\kern-0.2em%
 \raisebox{-0.3em}2}%
}
\newcommand\CETA{\systemname{C\kern-0.2exe\kern-0.5exT\kern-0.5exA}}
\newcommand{\DEDUCE}{D\scalebox{0.9}{EDUCE}\xspace}
\newcommand{\COMPOSE}{C\scalebox{0.9}{OMPOSE}\xspace}
\newcommand{\COLLAPSE}{C\scalebox{0.9}{OLLAPSE}\xspace}
\newcommand{\ORIENT}{O\scalebox{0.9}{RIENT}\xspace}
\newcommand{\DELETE}{D\scalebox{0.9}{ELETE}\xspace}
\newcommand{\SIMPLIFY}{S\scalebox{0.9}{IMPLIFY}\xspace}

\newcommand\irule[3][]{\frac{\strut\displaystyle#2}{\strut\displaystyle#3}_{#1}}
\newcommand\rstep{\to}

\newcommand\ro{>}
\newcommand\EE{\ensuremath{\mathcal{E}}\xspace}

\newcommand\RR{\ensuremath{\mathcal{R}}\xspace}
\newcommand{\from}{\leftarrow}
\def\test#1#2#3{\setbox0=\hbox{$\vphantom{#1}^{#2}_{#3}$}%
                \dimen0=\wd0%
                \setbox1=\hbox{$\scriptstyle #2$}%
                \advance\dimen0-\wd1%
                \setbox1=\hbox{\hskip\dimen0\copy1}%
                \dimen0=\wd0%
                \setbox2=\hbox{$\scriptstyle #3$}%
                \advance\dimen0-\wd2%
                \setbox2=\hbox{\hskip\dimen0\copy2}%
                {\vphantom{#1}^{\box1}_{\box2}}{#1}
}

\newcommand{\FromTBA}[3]{\mathrel{\test{#3}{#1}{#2}}}

\newcommand{\FromB}[1]{\FromTBA{}{#1}{\from}}

\begin{document}

\title{KBCV~{\cv} -- Automatic Completion\\ Experiments%
  \thanks{%
    Supported by the Austrian Science Fund (FWF) international project I963 and
    the Japan Society for the Promotion of Science.%
  }
}

\titlerunning{\KBCV~\cv}

\author{Thomas Sternagel}

\institute{
  University of Innsbruck,
  Innsbruck, Austria\\
  \email{thomas.sternagel@uibk.ac.at}
 }
\authorrunning{Sternagel}

\clearpage

\maketitle

\begin{abstract}
  This paper describes the automatic mode of the new version of the Knuth-Bendix
  Completion Visualizer. The internally used data structures have been
  overhauled and the performance was dramatically improved by introducing
  caching, parallelization, and term-indexing in the computation of critical
  pairs and simplification. The new version is much faster and can complete
  three more systems.
\end{abstract}



\section{Introduction}
\label{sec:intro}

The \emph{Knuth-Bendix Completion Visualizer} (\KBCV) is an
interactive/automatic tool for Knuth-Bendix completion and equational logic
proofs.
The basic functions of the previous release are described in detail
in~\cite{S12,SZ12}.
This paper addresses implementation issues to improve the performance of the
automatic completion mode and reports on experiments of the new release
\KBCV~{\cv}.
The tool is available under the \emph{GNU Lesser General Public License 3} at
\begin{center}
\url{http://cl-informatik.uibk.ac.at/software/kbcv}
\end{center}

In the sequel we assume familiarity with term rewriting, and
completion~\cite{BN98}. Nevertheless we recall the basics.

Completion is a procedure which takes as input a (finite) set of equations~$\EE$
and a reduction order $\ro$ (or it tries to construct this reduction order on
the fly with the help of an external termination tool, see~\cite{WSW06}) and
attempts to construct a terminating and confluent term rewrite system
(TRS)~$\RR$ with the same equational theory as~$\EE$.
In case the completion procedure succeeds, two terms are equivalent with respect
to~$\EE$ if and only if they reduce to the same normal form with respect
to~$\RR$, that is, $\RR$ represents a decision procedure for the word problem
of~$\EE$.

The computation is done by generating a finite sequence of intermediate TRSs
which constitute approximations of the equational theory of $\EE$. Following
Bachmair and Dershowitz~\cite{BD94} the completion procedure can be modeled as
an inference system (see Figure~\ref{fig:cir}).
The inference rules work on pairs $(\EE,\RR)$ where $\EE$ is a finite set of
equations and $\RR$ is a finite set of rewrite rules. The goal is to transform
an initial pair~$(\EE,\varnothing)$ into a pair $(\varnothing,\RR)$ such
that~$\RR$ is terminating, confluent and equivalent to~$\EE$. In our setting a
completion procedure based on these rules may succeed (find $\RR$ after finitely
many steps), loop, or fail.
In Figure~\ref{fig:cir} a reduction order~$\ro$ is provided as part of the
input.
We use $s\stackrel{\sqsupset}{\rstep}_\RR u$ to express that $s$ is reduced by a
rule $\ell\rstep r\in \RR$ such that $\ell$ cannot be reduced by another rule
with left-hand side $s$. The notation $\smash{s \stackrel{.}{\approx} t}$
denotes either of $s \approx t$ and $t \approx s$.

\KBCV internally uses indexed equations $i\colon l\approx r$ and rules $j\colon
l\to r$, where $i$ and $j$ are unique positive integers and $l$ and $r$ are
terms, called the left- and right-hand side respectively.

\begin{figure}[t]
  \begin{tabular}{@{ }r@{ }c@{ }l}
  \DEDUCE & $\irule{(\EE,\RR)}
  {(\EE\cup\{s\approx t\},\RR)}$ & if $s\FromB{\RR} u\rstep_\RR t$
  \\[3ex]
  \COMPOSE & $\irule{(\EE,\RR\cup\{s\rstep t\})}
  {(\EE,\RR\cup\{s\rstep u\})}$ & if $t\rstep_\RR u$
  \\[3ex]
  \COLLAPSE & $\irule{(\EE,\RR\cup\{s\rstep t\})}
  {(\EE\cup\{u\approx t\},\RR)}$ &
  if $s\stackrel{\sqsupset}{\rstep}_\RR u$
  \end{tabular}
  \hfill
  \begin{tabular}{@{ }r@{ }c@{ }l}
  \ORIENT & $\irule{(\EE\cup\{s\stackrel{.}{\approx}t\},\RR)}
  {(\EE,\RR\cup\{s\rstep t\})}$ & if $s \ro t$
  \\[3ex]
  \DELETE & $\irule{(\EE\cup\{s\approx s\},\RR)}
  {(\EE,\RR)}$
  \\[3ex]
  \SIMPLIFY & $\irule{(\EE\cup\{s\stackrel{.}{\approx}t\},\RR)}
  {(\EE\cup\{u\stackrel{.}{\approx}t\},\RR)}$ & if $s\rstep_\RR u$
  \end{tabular}
  \caption{The inference rules of \emph{completion}.}
  \label{fig:cir}
\end{figure}

\section{Optimizing Automatic Completion}
\label{sec:optac}

\KBCV 2.0 is implemented in Scala
2.10.0,\footnote{\url{http://www.scala-lang.org/}} an object-functional
programming language which compiles to Java bytecode. For this reason \KBCV is
portable and runs on Windows and Linux machines. The developed term library
({\tt scala-termlib}, available from \KBCV's homepage) was completely overhauled
and consists of approximately 2100 lines of code. The new \KBCV builds upon
this library and has an additional 5000 lines of code.

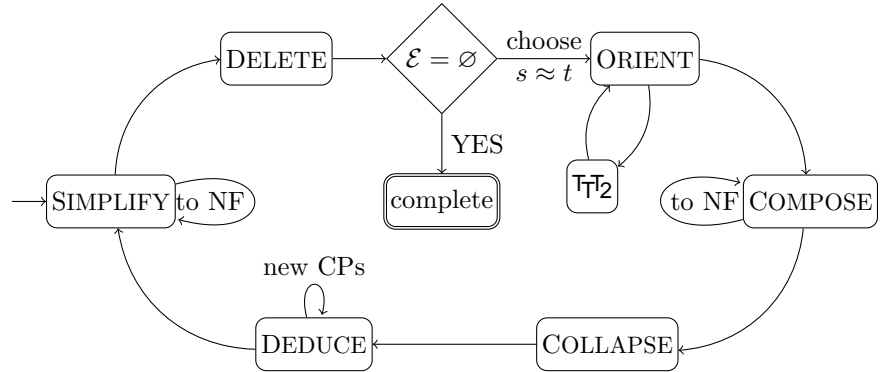
\begin{figure}[t]
\centering
\begin{tikzpicture}[node distance=2.2cm]
\tikzset{
    state/.style={
           rectangle,
           rounded corners,
           draw=black,
           minimum height=2em,
           inner sep=2pt,
           text centered,
           },
}
\tikzset{
    option/.style={
           diamond,
           draw=black,
           minimum height=2em,
           inner sep=0pt,
           text centered,
           },
}
  \node[initial by arrow,state,initial text=](1){\SIMPLIFY};
  \node[state,above of=1,right of=1,yshift=-3mm](2){\DELETE};
  \node[option,right of=2](3) {\begin{minipage}{1.2cm}\centering%
    $\EE=\varnothing$%
  \end{minipage}};
  \node[state,accepting,below of=3,yshift=3mm](4){complete};
  \node[state,right of=3,xshift=5mm](5){\ORIENT};
  \node[state,below of=5,right of=5,yshift=3mm](6){\COMPOSE};
  \node[state,below of=6,left of=6,xshift=-5mm,yshift=3mm](7){\COLLAPSE};
  \node[state,below of=1,right of=1,xshift=5mm,yshift=3mm](8){\DEDUCE};
  \node[state,left of=5,below of=5,xshift=15mm,yshift=5mm](9){\TTTT};
  \path[->](1) edge[bend left=40]  (2)
           (1) edge[loop right] node[anchor=east,left]{to NF} (1)
           (8) edge[loop above] node[anchor=north,above]{new CPs} (8)
           (2) edge node[above]  {} (3)
           (3) edge node[right]  {YES} (4)
           (3) edge node {\begin{tabular}{c}choose\\$s\approx t$\end{tabular}} (5)
           (5) edge[bend left=40]  (6)
           (6) edge[loop left] node[anchor=east,right]{to NF} (6)
           (6) edge[bend left=40]  (7)
           (7) edge node[above] {} (8)
           (8) edge[bend left=40]  (1)
           (5) edge[bend left=30] (9)
           (9) edge[bend left=30] (5)
  ;
\end{tikzpicture}
\caption{\KBCV's automatic completion procedure.}
\label{fig:iac}
\end{figure}

The main goal for this release was to improve the performance of \KBCV
especially in automatic mode. Some more details on the automatic mode can be
found in~\cite[Section 5.2.1]{S12} and~\cite[Section 2.2]{SZ12}.
Looking at the flow chart of the automatic mode depicted in Figure~\ref{fig:iac}
we first had to identify critical parts, where speed-up would be possible.

\begin{enumerate}
  \item The procedure starts in the \SIMPLIFY-phase, where both sides of
    equations are rewritten as far as possible.
  \item Trivial equations, that is, equations where
    both sides are the same are dropped in the \DELETE-phase.
  \item The third phase checks if $\EE$ is empty and if all critical pairs
    between left-hand sides of rules in $\RR$ are joinable.%
    \footnote{%
      The computation and check for joinability of critical pairs is only needed
      because of \KBCV's interactive mode in which inference rules may be fired
      in an arbitrary order.
    }
 \item Then the procedure chooses a single equation which it tries to orient.
   The used heuristic is to select an equation where the length of the left- and
   right-hand sides is minimal.
   The cost for orientation mainly depends on the used termination tool.
 \item Now in the \COMPOSE-phase the procedure simplifies all right-hand sides of
   rules as far as possible.
 \item After that, in the \COLLAPSE-phase, it tries to simplify left-hand sides
   of rules.
 \item Finally \DEDUCE computes critical pairs and adds them to the set of
   equations.
\end{enumerate}

From this assessment we see that (2) is trivial and already very fast and (4)
mainly depends on an external program. So we focus on the remaining phases.
In the sequel we will sometimes refer to (3) and (7) collectively as
\emph{critical pair computation} and to (1), (5), and (6) as
\emph{simplification}.

The first idea (which was already partly implemented in versions 1.7 and 1.8 of
\KBCV) was to prevent re-computation by introducing caching.
Next the independent parts of the procedure were parallelized to make the most
of modern multi-core/processor architectures.
Finally we also introduced term-indexing in order to speed up unification and
matching of terms.
These steps are described in some more detail in the next three sections.
We compare the resulting speed-ups for various combinations of these methods in
Section~\ref{sec:exper}.

\subsection{Caching}
\label{sec:cache}

In order to avoid redundancy in critical pair computations and simplifications
we introduced four new data structures for caching.

Each time a critical pair is computed \KBCV stores the pair of indices of the
overlapping rules which caused the new equation.
The next time automatic completion has to compute critical pairs (in phases (3)
or (7) of the procedure) it only computes critical pairs from overlaps which are
not already stored.

We use three different caches for \COMPOSE, \COLLAPSE, and \SIMPLIFY
respectively.
The first cache has an entry for each rule. In this entry we store the
set of indices of rules which have already been tried to simplify this rule.
Next time automatic completion has to simplify a rule it only tries the rules
which are not cached yet.
The other two caches work just in the same way.

\subsection{Parallelization}
\label{sec:paral}

While automatic completion (Figure~\ref{fig:iac}) executes the single phases
sequentially, within a phase there are completely independent computations which
can be parallelized.
\begin{itemize}
  \item\DEDUCE: The computation of critical pairs.
  \item\COMPOSE: The composition of rules.
  \item\COLLAPSE: The collapsing of rules.
  \item\SIMPLIFY: The simplification of equations.
\end{itemize}
In order to get the most out of modern multi-core architectures we
re-implemented those four phases. Now each single step (e.g.\ the computation of
critical pairs between two particular rules, or one specific rewrite step on one
side of an equation) are separate computations which can be handled by a pool of
worker threads. The main program waits until all results are computed and then
continues with the non-parallel part of the procedure.

\subsection{Term Indexing}
\label{sec:termi}

Both unification of terms (needed for the computation of critical pairs) and
matching (needed for rewriting of terms) can get very expensive for large
systems with large left-hand sides of rules.
To counteract that we now store the left-hand sides of rules in a discrimination
tree (see for example~\cite{RSV01}) which allows for very fast filtering of so
called \emph{candidate sets} (which are typically very small). Getting a
unifiable or matching term from this candidate set is much faster than checking
all left-hand sides of rules.

\section{Experiments}
\label{sec:exper}

\begin{table}[t]
\centering
\begin{tabular}{@{}l@{}c@{\ }c@{\ }c@{\ }c@{\ }c@{\ }c@{\ }c@{\ }c@{}}
  & {\tt KBCV-b-i-u}
  & {\tt KBCV-i-u}
  & {\tt KBCV-b-u}
  & {\tt KBCV-b-i}
  & {\tt KBCV-u}
  & {\tt KBCV-i}
  & {\tt KBCV-b}
  & {\tt KBCV}\\
\hline
\emph{completed} & 85 & 87 & 85 & 85 & 89 & 90 & 85 & 90 \\
\emph{total time} & 1142.6 & 512.8 & 498.0 & 384.5 & 1163.4 & 1321.3 & 321.8 & 1116.2 \\
\emph{avg. time} & 13.4 & 5.9 & 5.9 & 4.5 & 13.1 & 14.7 & 3.8 & 12.4 \\
\hline
{\tt AD93\_Z22} & & 83.9 & & & 44.7 & 41.8 & & 32.9 \\
{\tt BGK94\_D16} & & 30.5 & & & 25.7 & 25.6 & & 23.2 \\
{\tt BGK94\_Z22W} & & & & & 598.7 & 220.6 & & 201.6 \\
{\tt LS94\_G1} & & & & & & 583.6 & & 514.5 \\
{\tt SK90\_3.09} & & & & & 168.1 & 161.1 & & 90.1 \\
\hline
\end{tabular}
\caption{Experimental results on 115 systems, timeout: 600s.}
\label{tab:exp}
\end{table}

The experiments we describe here were carried out on a 64bit GNU/Linux machine
with 48 AMD Opteron{\texttrademark} 6174 processors and 315 GB of RAM.
The kernel version is 2.6.32.
The version of Java on this machine is 1.7.0\_03. For the JVM we limited the
stack size for each thread to 10MB, set the initial heap size to 1GB, and the
maximum heap size to 2GB.
The test-bed we worked with consists of 115 systems from the distribution of
\MKBTT.%
\footnote{\url{http://cl-informatik.uibk.ac.at/software/mkbtt/index.php}}
\KBCV was launched using the following flags: 
\begin{center}
  {\tt ./kbcv -a -p -s 600 -m "./ttt2 -cpf xml - 1" <inputfile>}
\end{center}
Here the {\tt -a} flag tells \KBCV to switch to automatic mode, {\tt -p} causes
\KBCV to output the CPF proof of completion on {\tt stdout}, {\tt -s 600} sets
the timeout to 600 seconds and finally {\tt -m} sets the termination-check
method to use, in our case calls to the external termination tool \TTTT.  There
are three more flags we used in the experiments: {\tt -b} disables caching, {\tt
-i} disables term-indexing, and {\tt -u} disables parallelization. The tool
instances where parallelization was enabled used all of the 48 processors.

The upper part of Table~\ref{tab:exp} gives the number of completed systems, the
total time needed to complete them and the average time for each of the
completed systems for different configurations of \KBCV.
The lower part lists systems which only certain configurations of \KBCV could
complete together with the time.
The detailed experiments are available online.%
\footnote{\url{http://cl-informatik.uibk.ac.at/software/kbcv/experiments/kbcv2/}}
Here each column is labeled with {\tt KBCV} plus the set flags. So the first
column labeled {\tt KBCV-b-i-u} gives the results for \KBCV without caching,
term-indexing, and parallelization, while the last column shows the results for
\KBCV using all three methods.
What we see is that without optimization \KBCV can complete 85 out of the 115
systems and the average time for that is 13.4 seconds per system.
Without caching (columns three, four, and seven) we are not able to complete
additional systems, although we achieve a speed-up of about 2.6 using only
term-indexing or parallelization, and 3.5 using both of these methods.
Only using caching (column two) already establishes two more systems with a
speed-up for the initial 85 systems of about 2.9.
Caching plus term-indexing (column five) already yields two more successful
systems and a speed-up with respect to the 85 systems of about 3.5.
If we combine caching with parallelization (column six) we get yet another
system and a speed-up for the initial systems of about 4.0.
Finally \KBCV using all three methods achieves a speed-up of 4.5 for the initial
85 systems.
All found proofs have been certified by \CETA~\cite{STZS12}.

\section{Conclusion}
\label{sec:concl}

Three different methods to enhance \KBCV's automatic completion procedure have
been investigated and compared.
We have seen that these methods, most notably caching, achieve a huge
performance boost for the automatic completion procedure of \KBCV~{\cv}.

If we look at the 115 systems we tested, we see that most of them only consist
of about 10 to 20 rules and that the left-hand sides of those are also pretty
small. When we work with much larger systems with more complicated left-hand
sides parallelization and term-indexing become more and more important.
We for example tried to only compute critical pairs for a subset of the
\HOLLight~\cite{H96} simpset (about 3000) rules. Without parallelization we had
to cancel the experiment after several days. Using parallelization \KBCV was
able to compute the 300,000 critical pairs in less than two hours.

A next step to further push the procedure would be to investigate different
heuristics for the selection of equations in the \ORIENT-phase.

%
\label{sect:bib}
\bibliographystyle{plain}
\bibliography{references}

\end{document}